\documentclass[aps,pra,twocolumn,showpacs]{revtex4}
\usepackage{graphicx}
\usepackage{dcolumn}
\usepackage{amsmath}

\begin{document}

\title{Stability diagram and growth rate of parametric resonances \\
in Bose-Einstein condensates in one-dimensional optical lattices}

\author{C. Tozzo$^1$, M.~Kr\"amer$^2$, and F. Dalfovo$^1$}
\affiliation{$^1$BEC-INFM and Dipartimento di Fisica, Universit\`a
di Trento, I-38050 Povo, Italy \\
$^2$ JILA, University of Colorado and \\ National Institute of
Standards and Technology, Boulder, CO 80309-0440, USA }

\date{May 24, 2005}

\begin{abstract}
A Bose-Einstein condensate in an optical lattice exhibits parametric
resonances when the intensity of the lattice is periodically modulated
in time. These resonances correspond to an exponential growth of the
population of counter-propagating Bogoliubov excitations. A suitable 
linearization of the Gross-Pitaevskii (GP) equation is used to calculate
the stability diagram and the growth rates of the unstable modes. The 
results agree with the ones extracted from time-dependent GP simulations,
supporting our previous claim (M. Kr\"amer {\it et al.}, Phys. Rev. A 
(2005) in press) concerning the key role of parametric resonances in the 
response observed by St\"oferle {\it et al.} (Phys. Rev. Lett. {\bf 92}, 
130403 (2004)) in the superfluid regime. The role of the seed excitations 
required to trigger the parametric amplification is discussed. The possible 
amplification of the quantum fluctuations present in the quasiparticle 
vacuum, beyond GP theory, is also addressed, finding interesting analogies 
with similar processes in nonlinear quantum optics and with the dynamic 
Casimir effect. Our results can be used in exploiting parametric 
instabilities for the purpose of spectroscopy, selective amplification 
of a particular excitation mode and for establishing a new type of 
thermometry.
\end{abstract}


\maketitle

\section{Introduction}

A parametric resonance corresponds to the exponential growth of
certain modes of a system induced by the periodic variation of a
parameter \cite{landau,arnold}. It is a very general phenomenon 
occurring in classical oscillators, in nonlinear optics, in systems 
governed by Non-Linear Schr\"odinger equations, and in Hamiltonian 
chaotic systems. Parametric resonances can also occur in 
Bose-Einstein condensates made of ultracold atomic gases. From the 
theoretical side, they have already been investigated in 
3D \cite{castin,kagan,kevrekidis} and 2D \cite{ripoll,staliunas} 
harmonically trapped condensates, for condensates in oscillating 
double-well potentials \cite{salasnich,salmond,harout} and in deep 
optical lattices \cite{rapti}. These investigations are based 
on suitable effective Hamiltonians in which the interaction between 
the atoms enters through a mean-field term proportional to the 
condensate density. Most of them make use of the Gross-Pitaevskii 
(GP) equation, which has the form of a Non-Linear Schr\"odinger 
equation and gives a very accurate description of dilute 
condensates \cite{rmp,book}.  

In Ref.~\cite{kraemer} we have recently shown that parametric
resonances can be the origin of the large and broad response observed
by Esslinger and co-workers \cite{stoeferle,koehl,schori} in the
superfluid phase of a condensate in an optical lattice. Our claim
was based on the results of numerical simulations, namely the
integration of the time-dependent GP equation. In the present
work we perform a more systematic analysis, by investigating the
behavior of Bogoliubov quasiparticles subject to a periodic modulation
of the lattice. The purpose is to gain a deeper insight into the
process of parametric resonances, in particular into the conditions
for the occurrence of the instability, the growth rates associated
with the parametric amplification of an excitation and the role of
the initial fluctuations (seed excitations) which trigger the onset
of the amplification.

The basic theory is presented in section \ref{section-bogo}. We 
consider an infinite condensate at zero temperature, subject to 
a transverse harmonic potential and to a 1D optical lattice 
along the $z$-direction, with $N$ atoms per lattice site. The 
effect of the optical lattice can be expressed through the periodic 
potential $V (z,t) = s(t) E_R \sin^2 (q_B z)$, where $q_B=\pi/d$ 
is the Bragg wave vector, $d$ is the lattice spacing and $s$ is 
the lattice depth in units of the recoil energy $E_R= \hbar \omega_R 
= \hbar^2 q_B^2/2m$. The transverse harmonic confinement, of
frequency $\omega_\perp$, is taken to be sufficiently strong to
inhibit excitations involving the radial degrees of freedom on the
energy scale we are concerned with. Under this assumption, the
order parameter can be factorized into a radial Gaussian, having
constant width $a_{\perp}= [\hbar/(m\omega_{\perp})]^{1/2}$, times
a $z$ and $t$-dependent order parameter, $\Psi(z,t)$. The latter
obeys an effective 1D GP equation which includes the
atom-atom interaction through a coupling constant $g$. We first
consider the case of a static optical lattice ($s(t)=s_0$) and
we find the Bogoliubov spectrum as a result of the linearization
of the GP equation with regard to a small change, $\delta\Psi(z,t)$, of
the condensate wave function. Then we study the evolution of the
system under a modulation of lattice depth in the form $s (t) =
s_0 [1 + A \sin(\Omega t)]$. A stability analysis is done in order 
to find the parametrically unstable modes and calculate their growth
rate for given values of the parameters $s_0$, $A$ and $gn$,
where $n=N/d$ is the average linear density. We use a Bogoliubov
quasiparticle projection method \cite{ballagh,tozzo} in order to
provide a deeper insight into the mechanisms of instability.

When $A$ is not too large, the main mechanism which drives 
a parametric resonance is a coupling between pairs of
counter-propagating Bogoliubov axial excitations of frequencies
$\omega_{jk}$ and $\omega_{j'k}$, where $k$ is the wave vector and
$j$ and $j'$ are Bloch band indexes. The multi-mode stability 
analysis then reduces to a two-mode approximation. The resonance 
condition becomes $\Omega = \omega_{jk} + \omega_{j'k}$, which 
yields $\Omega = 2 \omega_{0k}$ for axial phonons in the lowest 
Bloch band. This approximation is discussed in section 
\ref{section-2mode},  where we show that the parameters $s_0$, 
$A$ and $gn$ can be combined in a single parameter 
$\overline{\gamma}= (1/2) s_0 A \Omega \Gamma^{jj'}_{+-}$, which 
contains all relevant information about the stability diagram 
and the growth rate. The quantity $\Gamma^{jj'}_{+-}$ depends on 
$s_0$ and $k$ and can be calculated by solving the Bogoliubov 
equations for different lattice depths. It describes the coupling 
between Bogoliubov modes belonging to the $j$ and $j'$ bands and 
having opposite quasi-momenta. This coupling is mediated by the 
time-dependence of the lattice potential. In the tight binding 
regime, i.e. when the lattice depth is large enough to split up 
the condensate in almost independent condensates at each lattice 
site, a semi-analytic result can be derived for the quantity 
$\overline{\gamma}$ describing the parametric resonance of modes
of the lowest Bogoliubov band. In this case, in fact, the
quantity $\Gamma^{00}_{+-}$ can be simply expressed through the
$s$-dependence of the effective mass and the compressibility of
the condensate. This regime is discussed in section
\ref{section-tb}.

The numerical results of the stability analysis are shown in section
\ref{section-results} together with those of the two-mode and
tight-binding approximations. The two-mode approximation turns out
to be very accurate in the whole range of parameters here considered.
At large $s_0$ the tight binding expressions work well. The approach
to this regime is faster when the mean-field interaction is small
(i.e., small values of $gn$).

In section \ref{section-gp} the results obtained in the
previous sections by linearizing the GP equation for an infinite 
condensate are compared with those obtained
from the numerical integration of the time dependent GP equation.
We perform GP simulations for an infinite condensate,  as well as
a trapped condensate similar to the ones of the experiments of
Refs.~\cite{stoeferle,koehl,schori} and of our previous calculations
of Ref.~\cite{kraemer}. The agreement is good. This confirms the
overall picture for the key role played by the parametric instability
of Bogoliubov excitations in the observed large and broad response
of the condensate to the modulation of the lattice depth.  These
results also support the possibility to use parametric resonances
as a tool for spectroscopic studies of Bogoliubov excitations, for 
a selective amplification of certain modes and for the characterization 
of thermal and/or quantum fluctuations in
trapped condensates. A brief discussion on how parametric resonances
can give information about thermal and quantum fluctuations is given
in section \ref{section-seed}.

%%%%%%%%%%%%%%%%%%%%%%%%%%%%%%%%%%%%%%%%%%%%%%%%%%%%%%%%%%%%%%%%%%%%%%%%%%%
\section{Bogoliubov excitations in an amplitude modulated lattice}
\label{section-bogo}

%%%%%%%%%%%%%%%%%%%%%%%%%%%%%%%%%%%%%%%%%%%%%%%%%%%%%%%%%%%%%%%%%%%%%%%%%%%

\subsection{Bogoliubov spectrum in a static lattice}

Let us consider a dilute condensate made of atoms with mass $m$ and 
$s$-wave scattering length $a$, whose order parameter obeys the
time-dependent GP equation \cite{rmp,book}
\begin{equation}
i\hbar \partial_t \Psi =
\left[ -\frac{\hbar^2}{2m} \nabla^2
+ V +g_{3D} N |\Psi|^2\right] \Psi \, ,
\label{tdgpe}
\end{equation}
where $\Psi$ is normalized to $1$, $N$ is the number of particles and 
$g_{3D}=4\pi\hbar^2 a/m$. The potential $V$ is the sum of the harmonic 
trap and the optical lattice. The 1D optical lattice is taken to be 
oriented along the $z$-direction with lattice depth $s$, spacing $d$, 
Bragg wave vector $q_B=\pi/d$ and recoil energy $E_R= \hbar \omega_R =
\hbar^2 q_B^2/2m$. The transverse harmonic confinement is assumed to be 
sufficiently strong to freeze the radial degrees of freedom, so that 
the order parameter is simply a Gaussian in the transverse direction, 
having constant width $a_{\perp}=[\hbar/(m\omega_{\perp})]^{1/2}$.  
Under this condition, the 3D GP equation (\ref{tdgpe}) reduces to an 
effective 1D GP equation for a purely axial order parameter $\Psi(z,t)$: 
\begin{equation}
i\hbar\partial_t\Psi=\left[-\frac{\hbar^2}{2m}
\partial_z^2+s E_R\sin^2(q_Bz) +gnd \ |\Psi|^2\right]\Psi \, .
\label{eq:TDGP}
\end{equation}
Here the normalization condition is $\int_{-d/2}^{d/2} dz |\Psi|^2 = 
1$, while $nd=N$ is the number of atoms per lattice site and $n$ is 
the average linear density. The quantity $g=g_{3D}/2\pi a_{\perp}^2$ 
is an effective 1D coupling constant accounting for the atom-atom 
interaction. With this choice, Eq.~(\ref{eq:TDGP}) also coincides 
with that of Ref.~\cite{epjd} for a uniform condensate in a 1D 
lattice. 

When the condensate is weakly perturbed, the order parameter can be
written as
\begin{equation}
\Psi (z,t) = e^{-i\mu t/\hbar} [\Psi_0 (z) + \delta \Psi (z,t)]
\label{deltapsi1}
\end{equation}
where $\Psi_0$ is the groundstate solution of the stationary GP 
equation
\begin{equation}
\left[-\frac{\hbar^2}{2m} \partial_z^2+s E_R\sin^2(q_Bz) +gnd
\ |\Psi_0|^2\right]\Psi_0  = \mu \Psi_0
\label{eq:stationaryGP}
\end{equation}
and
$\mu$ is the corresponding chemical potential. The small variation
$\delta\Psi$ can be expressed in terms of Bogoliubov excitations,
\begin{equation}
\delta\Psi=\sum_{jk} c_{jk}  u_{jk}(z) e^{-i\omega_{jk}t}
+c_{jk}^*  v_{jk}^*(z) e^{i\omega_{jk}t} \, ,
\label{deltapsi2}
\end{equation}
where $c_{jk}$ are complex  coefficients. Due to the periodicity 
of the external potential, the Bogoliubov quasiparticle amplitudes 
$u_{jk}$ and $v_{jk}$ have the form of Bloch functions $u_{jk}=
\exp(ikz) \tilde{u}_{jk}$ and $v_{jk}=\exp(ikz)\tilde{v}_{jk}$, where
$\tilde{u}_{jk}$ and $\tilde{v}_{jk}$ are periodic with period
$d$. The excitations are labelled by their quasimomentum $\hbar k$
and the Bloch band index $j$.

By inserting (\ref{deltapsi2}) into (\ref{eq:TDGP}) and linearizing
with respect to $\delta\Psi$ one finds that the amplitudes $u_{jk}$ and
$v_{jk}$ are solutions of the eigenvalue problem
\begin{equation}
{\mathcal L}(s,k)\left(
\begin{array}{c}\tilde u_{jk}\\ \tilde v_{jk}\end{array}
\right) =\hbar\omega_{jk}\left(
\begin{array}{c}\tilde u_{jk}\\ \tilde v_{jk}\end{array}\right)
\label{eq:BOGO}
\end{equation}
where the matrix ${\mathcal L}(s,k)$ is given by
\begin{eqnarray}
{\mathcal L}(s,k)=
\left(\begin{array}{cc}H(s,k)&gnd\Psi_0^2\\
-gnd\Psi_0^{*2}&-H(s,k)\end{array} \right)
\label{eq:Ldefinition}
\end{eqnarray}
and we have defined
\begin{eqnarray}
H(s,k) &=& -\frac{\hbar^2}{2m}(\partial_z+ik)^2+sE_R\sin^2(q_Bz)
\nonumber \\
&+& 2gnd \ |\Psi_0|^2-\mu\;.
\label{eq:HGP}
\end{eqnarray}
Moreover, for any given $k$ the solutions $\tilde u_{jk}$ and 
$\tilde v_{jk}$  obey the ortho-normalization conditions
\begin{eqnarray}
\int_{-d/2}^{d/2} \!dz \, ( \tilde u_{jk}^* \tilde u_{j'k}
  - \tilde v_{jk}^* \tilde v_{j'k} )&=& \delta_{jj'}\,,
\label{eq:uvnorm1}\\
\int_{-d/2}^{d/2} \!dz \, ( \tilde u_{jk} \tilde v_{j',-k}
  -  \tilde u_{j',-k} \tilde v_{jk} ) &=& 0\,.
\label{eq:uvnorm2}
\end{eqnarray}

\begin{figure}[h]
\begin{center}
\includegraphics[width=8cm]{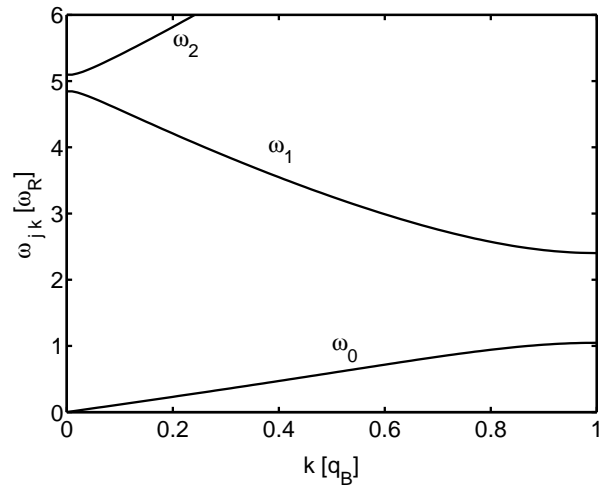}
\end{center}
\caption{Lowest bands in the Bogoliubov spectrum of a cylindrical 
condensate in a periodic lattice, obeying Eq.~(\protect\ref{eq:TDGP}) 
with $s=s_0=4$ and $gn=0.72 E_R$. The range of $k$ corresponds to 
half of the first Brillouin zone. We use the shorthand notation 
$\omega_0$, $\omega_1$ and $\omega_2$ to indicate the Bogoliubov 
bands whose dispersion relation is $\omega_{0k}$, $\omega_{1k}$ 
and $\omega_{2k}$. }
\label{QP0}
\end{figure}

Equations (\ref{eq:BOGO}) can be solved numerically to get the amplitudes 
$\tilde u_{jk}$, $\tilde v_{jk}$ and the Bogoliubov band spectrum
$\hbar\omega_{jk}$ for given values of the lattice depth $s=s_0$ and
interaction parameter $gn$.  In Fig.~\ref{QP0} we plot the lowest three
bands  as obtained for $s_0=4$
and $gn=0.72 E_R$. The latter value is chosen as the axial
average of the interaction parameter in the experimental setting of
\cite{stoeferle}. The lowest band exhibits a linear dispersion for
$k \to 0$ as expected for long wavelength Bogoliubov phonons traveling
along $z$ at the speed of sound. This type of spectra have been already
discussed in the recent literature \cite{berg,chiofalo,epjd}. Here
we use the results of Fig.~\ref{QP0} as a basis for the subsequent
discussion of parametric resonances.

%%%%%%%%%%%%%%
%%%%%%%%%%%%%%

%%%%%%%%%%%%%%%%%%%%%%%%%%%%%%%%%%%%%%%%%%%%%%%%%%%%%%%%%%%%%%%%%%%%%%%%%%%
\subsection{Modulation of the lattice depth}
\label{section-mod}

Let us consider a periodic modulation of the lattice depth in the
form
\begin{equation}
s (t) = s_0 [1 + A \sin(\Omega t)] \; ,
\label{eq:modulation}
\end{equation}
with $A \ll 1$. The dynamics induced by this modulation can still
be described by Eqs.~(\ref{eq:TDGP}) and (\ref{deltapsi1}). To
this purpose we assume that $\Psi_0$ is, at any time $t$, the
stationary solution of (\ref{eq:stationaryGP}) for $s=s(t)$. Both
$\Psi_0$ and $\mu$ are now functions of $t$ {\it via} their dependence 
on $s(t)$. Then we insert again
(\ref{deltapsi1}) into (\ref{eq:TDGP}), take $s$ as in
(\ref{eq:modulation}), and look for an equation of motion for the
small part $\delta \Psi$. Up to linear terms in $\delta \Psi$, one
gets
\begin{eqnarray}
i \partial_t \delta \Psi &=&
[ H(s,k) + 2gnd \ |\Psi_0|^2 - \mu] \delta \Psi
\nonumber \\
&+& gnd \Psi_0^2 \delta \Psi^* + \dot{s} [ i \partial_s
+ (\partial_s \mu)t/\hbar] \Psi_0 \; .
\label{newTDGP}
\end{eqnarray}
The inhomogeneous term, proportional to $\dot{s}$, behaves like a
source of excitations in the linear response regime. It creates
excitations out of the condensate when $\Omega$ is resonant with a
Bogoliubov frequency $\omega_{jk}$ at $k=0$. For symmetry reasons,
the first resonance occurs at the bottom of the $\omega_2$ band.
In the case of Fig.~\ref{QP0} this amounts to $\Omega=
\omega_2(k=0) \simeq 5.1 E_R /\hbar$. Similar resonances can be
found at larger frequencies. If $A$ is small and the modulation time
is long, this type of excitation processes only occurs in narrow
and well separated intervals of $\Omega$.  Except in these
intervals, the inhomogeneous term in Eq.~(\ref{newTDGP}) can be
safely ignored. In the range of $\Omega$  of the two lowest Bogoliubov
bands $\omega_0$ and $\omega_1$, for instance, this approximation
is expected to work well, as confirmed also by the comparison with
GP simulations (see section \ref{section-gp} below). In the same
range, also the assumption that $\Psi_0$ adiabatically follows the
lattice modulation is expected to be valid. Finally, the
periodicity of the system allows one to write the unknown function
$\delta \Psi (z,t)$ in the form of a Bloch wave expansion
\begin{equation}
\delta\Psi(z,t)= \sum_{k>0}
\tilde a_{+k}(z,t) e^{ikz}+ \tilde a_{-k}^*(z,t)e^{-ikz}\,,
\label{eq:uvansatz}
\end{equation}
where the coefficients $\tilde a_{k}(z,t)$ are periodic in $z$ with
period $d$. Neglecting the last term in (\ref{newTDGP}) and inserting
the {\it ansatz} (\ref{eq:uvansatz}), one finally gets
\begin{equation}
i\hbar\partial_t
\left(\begin{array}{c} \tilde a_{+k} \\ \tilde a_{-k}\end{array} \right)=
{\mathcal L}(s,k)
\left(\begin{array}{c} \tilde a_{+k} \\ \tilde a_{-k}\end{array} \right)\,.
\label{eq:rescaledBOGO}
\end{equation}
This is the equation of motion for $\delta \Psi$. It contains the
external parameter $s(t)$, which varies periodically in time, and
hence it can exhibit parametric instabilities. We perform the stability
analysis of this equation using the method explained in section
\ref{section-multimode} below.

In order to give a more transparent view of the processes involved
in Eq.~(\ref{eq:rescaledBOGO}), one can use a suitable basis of
Bogoliubov states \cite{ballagh,tozzo}.  Let us write
\begin{equation}
\left(
\begin{array}{c}
\tilde a_{+k}\\ \tilde a_{-k}
\end{array}
\right)
=
\sum_j\left[
c_{jk}
\left(
\begin{array}{c}
\tilde u_{jk}\\ \tilde v_{jk}
\end{array}
\right)
+
c_{j,-k}^*
\left(
\begin{array}{c}
\tilde v_{j,-k}^*\\ \tilde u_{j,-k}^*
\end{array}
\right)
\right]\,
\label{atilde}
\end{equation}
where $\tilde u_{jk}$, $\tilde v_{jk}$ denote the solutions of 
(\ref{eq:BOGO}) at the lattice depth $s(t)$. This corresponds to the 
projection of $\delta\Psi$, at time $t$, onto the instantaneous 
Bogoliubov modes at lattice depth $s(t)$. Inserting (\ref{atilde}) into
(\ref{eq:rescaledBOGO}) and using (\ref{eq:uvnorm1})-(\ref{eq:uvnorm2}), 
one finds the following equations for the time-dependent 
coefficients $c_{jk}$:
\begin{equation}
i\partial_t c_{jk} = \omega_{jk} c_{jk} -i\dot{s}\sum_{j'} \left[
  \Gamma_{++}^{jj'} c_{j'k} + \Gamma_{+-}^{jj'} c_{j',-k}^*\right]
\label{c_jk}
\end{equation}
\begin{equation}
i\partial_t c_{j,-k}^* = - \omega_{jk} c_{j,-k}^* -i\dot{s}\sum_{j'}
\left[ \Gamma_{+-}^{jj'} c_{j'k} + \Gamma_{++}^{jj'}
  c_{j',-k}^*\right]\,,
\label{c_j-k}
\end{equation}
where
\begin{eqnarray}
\Gamma_{++}^{jj'}&=&\int_{-d/2}^{d/2} ~\hskip -0.5cm dz
\left(\tilde u_{jk}^*\partial_s
\tilde u_{j'k}- \tilde v_{jk}^*\partial_s \tilde v_{j'k}\right)\,,
\label{eq:Gamma++jj'}\\
\Gamma_{+-}^{jj'}(s)&=&\int_{-d/2}^{d/2} ~\hskip -0.5cm dz
\left(\tilde u_{jk}^*\partial_s
\tilde v_{j',-k}^*- \tilde v_{jk}^*\partial_s \tilde
u_{j',-k}^*\right).
\label{eq:Gamma+-jj'}
\end{eqnarray}
In deriving (\ref{c_jk})-(\ref{c_j-k}) we have used the relations 
$\tilde u_{jk}=\tilde u_{j,-k}^*$ and $\tilde v_{jk}=\tilde v_{j,-k}^*$, 
which follows from the fact that the groundstate wavefunction can be 
chosen real (zero current).

Eqs.~(\ref{c_jk})-(\ref{c_j-k}) reveal that the time-dependence
of the lattice depth ($\dot{s}\neq 0$) brings about a coupling
between Bogoliubov excitations with the same $|k|$. In particular,
the quantity $\Gamma_{++}^{jj'}$ accounts for the scattering of
excitations from the perturbation, that is a process in which an
excitation in the band $j$ is scattered into the band $j'$ without
changing its quasimomentum $k$. This happens resonantly when $\Omega$ 
is equal to the frequency difference between two bands at the same $k$.
One can easily prove that this process does not vary the total 
number of excitations \cite{notegamma++}. Conversely, the quantity 
$\Gamma_{+-}^{jj'}$ 
describes the coupling between counter-propagating excitations.
As we will show in the following, this quantity causes an exponential
growth of the number of excitations and is thus responsible for 
the parametric instability. It is worth noticing that 
$\Gamma_{+-}^{jj'}=0$ if $gn=0$ and hence a noninteracting gas, 
obeying the Schr\"odinger equation, does not exhibit parametric 
resonances.

%%%%%%%%%%%%%%%%%%%%%%%%%%%%%%%%%%%%%%%%%%%%%%%%%%%%%%%%%%%%%%%%%%%%%%%%%%
\subsection{Stability analysis}
\label{section-multimode}

In order to find the parametrically unstable solutions of the linear
equation (\ref{eq:rescaledBOGO}) we follow the procedure discussed in
Ref.~\cite{shirts}. The dynamics is determined by the operator
${\mathcal L}$, which changes in time {\it via} its dependence on
$s(t)$. The lattice modulation is periodic with period $T=2\pi/\Omega$.
Therefore the operator ${\mathcal L}$ has the same periodicity and can
be expressed by means of the Fourier representation
\begin{equation}
{\mathcal L}(t)=\sum_{\nu=-\infty}^{+\infty}{\mathcal L}^{(\nu)}
e^{i \nu \, \Omega t}\;, \label{eq:Lfourier}
\end{equation}
where $\nu$ is integer and the operators ${\mathcal L}^{(\nu)}$ do
not depend on time. Thanks to Floquet theorem \cite{floquet}, any
solution of (\ref{eq:rescaledBOGO}) can be written in the form
\begin{equation}
\left(\begin{array}{c} \tilde a_{+k}(t) \\ \tilde a_{-k}(t)
\end{array} \right)=
e^{\lambda t } \left(\begin{array}{c}  \tilde{a'}_{+k}(t) \\
\tilde{a'}_{-k}(t)\end{array} \right)\; ,
\label{eq:floquet}
\end{equation}
where $\lambda$ is a complex constant and the functions
$\tilde{a'}_{\pm k}(t)$ are periodic of period $T$. This periodicity
allows one to expand $\tilde{a'}_{\pm k}$ in the Fourier series
\begin{equation}
\tilde{a'}_{\pm k}(t)=\sum_{\nu=-\infty}^{+\infty} \tilde
a^{(\nu)}_{\pm k} e^{i \nu \, \Omega t}\;. 
\label{eq:afourier}
\end{equation}
Finally, by using Eqs.~(\ref{eq:Lfourier})-(\ref{eq:afourier}) one
can rewrite Eq.~(\ref{eq:rescaledBOGO}) in the form
\begin{equation}
i \lambda \left(\begin{array}{c} \tilde a^{(\nu)}_{+k} \\
\tilde a^{(\nu)}_{-k}\end{array} \right) =
\nu \hbar \Omega \left(\begin{array}{c} \tilde a^{(\nu)}_{+k} \\
\tilde a^{(\nu)}_{-k}\end{array} \right) + \! \sum^{+\infty}_{\mu
=-\infty} {\mathcal L}^{(\mu-\nu)} \left(\begin{array}{c} \tilde
a^{(\mu)}_{+k} \\ \tilde a^{(\mu)}_{-k}
\end{array} \right) .
\label{eq:stabilitydiagonalization}
\end{equation}
 From the solutions of this eigenvalue problem one obtains the 
complex eigenvalues $\lambda$ as a function of $k$ and $\Omega$.  
In practice, the solution can be found by truncating the series
(\ref{eq:afourier}) at a certain $\nu$ so that the problem is
reduced to a matrix diagonalization. In our case, we obtain a
very good convergence by cutting at $\nu$ of the order of
$10$. 

We then define the growth rate $\gamma$ of a Bogoliubov mode as the
real part of $\lambda$:
\begin{equation}
\gamma={\rm Re}(\lambda)\;.
\label{relambda}
\end{equation}
This quantity fixes the character of the corresponding
eigenfunction. If $\gamma\leq 0$ then $|\tilde a_{\pm k}(t)|$ is
always bound and we say that the mode $\pm k$ is parametrically
stable. On the contrary, if $\gamma>0$ the corresponding
eigenfunction describes an unstable mode whose population
exponentially increases in time. For a given modulation frequency
$\Omega$ we say that the condensate is parametrically stable if
all eigenfunctions of (\ref{eq:stabilitydiagonalization}) are
stable. By using symmetry properties analogous to the
ortho-normality relations (\ref{eq:uvnorm1}) and
(\ref{eq:uvnorm2}), one can prove that the condensate is
parametrically stable if and only if $\gamma=0$ for all modes. The
results of this stability analysis will be given in section
\ref{section-results}.

%%%%%%%%%%%%%%%%%%%%%%%%%%%%%%%%%%%%%%%%%%%%%%%%%%%%%%%%%%%%%%%%%%%%%%%%%
\section{Two-mode approximation}
\label{section-2mode}

According to Eqs. (\ref{c_jk})-(\ref{c_j-k}), the evolution of a
certain mode is, in principle, coupled to the modes of all bands
with the same $|k|$. However, one might expect that each resonance
essentially arises from the coupling of a pair of counter-propagating
modes, the influence of all the others being negligible.
In our formalism, this is equivalent to replacing the sum 
over $j'$ in Eqs.~(\ref{c_jk})-(\ref{c_j-k}) with a single term. 
The problem is thus reduced to the solution of the
coupled equations
\begin{eqnarray}
&& i\partial_t \left(\begin{array}{c} c_{jk}
\\ c_{j',-k}^*\end{array} \right)=
\nonumber\\
&&
\left(\begin{array}{cc}
                      \omega_{jk} - i \dot{s} \Gamma_{++}^{jj'} &
                      -i \dot{s} \Gamma_{+-}^{jj'}\\
                      -i \dot{s} \Gamma_{+-}^{jj'}&
                      -\omega_{j'k} - i \dot{s} \Gamma_{++}^{jj'}
\end{array}\right)
\left(\begin{array}{c} c_{jk} \\ c_{j',-k}^*\end{array} \right)
\label{eq:c+-}
\end{eqnarray}
for the coefficients $c_{jk}$ and $c_{j',-k}$ of the $\pm k$-modes in the
band $j$ and $j'$ respectively. The coupling $\Gamma_{+-}^{jj'}$ is the
same already given in (\ref{eq:Gamma+-jj'}).

The modulation (\ref{eq:modulation}) of the lattice explicitly enters
the two-mode equation (\ref{eq:c+-}) through the term $\dot{s}=A s_0
\Omega\cos(\Omega t)$ in the matrix elements. An additional implicit
dependence arises from the $s$-dependence of $\omega_{jk},\omega_{j'k}$,
$\Gamma_{++}^{jj'}$ and $\Gamma_{+-}^{jj'}$. In order to investigate
the regime of small modulation amplitude, one can linearize
Eq.~(\ref{eq:c+-}) with respect to $A$, keeping only the leading order
of each matrix element. One obtains
\begin{eqnarray}
&& i\partial_t  \left(\begin{array}{c} c_{jk}
\\ c_{j',-k}^*\end{array} \right) =
\nonumber\\
&&
\left(\begin{array}{cc} \omega_{jk}&
                      -i 2\overline{\gamma}\cos(\Omega t)\\
                      -i 2\overline{\gamma}\cos(\Omega t)&
                      - \omega_{j'k}\end{array}\right)
\left(\begin{array}{c} c_{jk} \\ c_{j',-k}^*\end{array} \right)
\label{eq:c+-approx}
\end{eqnarray}
where
\begin{equation}
\overline{\gamma}=\frac{1}{2} A s_0\Omega(\Gamma_{+-}^{jj'})_{s=s_0}
\label{gammamax2M}
\end{equation}
and
\begin{equation}
\omega_{jk}= (\omega_{jk})_{s=s_0} \; .
\end{equation}
A further simplification is obtained by neglecting the coupling
between resonances at $\pm\Omega$, which is justified when the
modulation time is much longer than $\Omega^{-1}$. This
allows one to replace $2 \cos(\Omega t)$ with $\exp(i\Omega t)$,
so that Eq.~(\ref{eq:c+-approx}) yields the two coupled equations
\begin{eqnarray}
\partial_t  f_{jk} &=& -\overline{\gamma} e^{i(\omega_{j'k}+
                                 \omega_{jk}-\Omega) t} f_{j',-k}^*
\label{fij}\\
\partial_t  f_{j',-k} &=& -\overline{\gamma} e^{i(\omega_{j'k}+
                                 \omega_{jk}-\Omega)t} f_{jk}^*
\label{eq:c+-approx2}
\end{eqnarray}
where we have introduced the new variable $f_{jk}=e^{i
\omega_{jk} t}c_{jk}$. By inserting (\ref{eq:c+-approx2}) in
the time derivative of (\ref{fij}), one gets
\begin{equation}
\left[\partial_t^2-i(\omega_{jk}+\omega_{j'k}-\Omega)
\partial_t-|\overline\gamma|^2\right]f_{jk}(t)=0
\label{eq:c+-1eq}
\end{equation}
with initial conditions
\begin{equation}
f_{jk}(0) = c_{jk}(0)
\; \; \; ; \; \; \;
\left(\partial_t f_{jk}\right)_{t=0} = -\overline\gamma c_{j',-k}^*(0)\;.
\label{eq:ci}
\end{equation}
One can easily see that the solutions of Eq.~(\ref{eq:c+-1eq})
exhibit an exponential growth within the unstable region
$|\omega_{jk}+\omega_{j'k} -\Omega| < 2|\overline{\gamma}|$, where 
the growth rate is found to be
\begin{equation}
\gamma= \frac{1}{2} [\, 4 \left|\overline{\gamma}\right|^2
-\left(\omega_{jk}+\omega_{j'k}- \Omega\right)^2 ]^{1/2}\, .
\label{gamma2Mapp}
\end{equation}
This also implies that the maximal growth occurs at the rate
\begin{equation}
\gamma_{\rm max}=\left|\overline{\gamma}\right|
\label{gammamax}
\end{equation}
when the resonance condition
\begin{equation}
\Omega = \omega_{jk} + \omega_{j'k}
\label{eq:res-cond}
\end{equation}
is met. The width of the unstable region
is $2 |\overline{\gamma}|$. By recalling definition (\ref{gammamax2M}),
one sees that both the maximum rate and the width of the unstable
region are proportional to the modulation amplitude $A$. We
notice that, when the coupling occurs between counter-propagating
modes in the same Bloch band ($j=j'$), the resonance condition
is simply $\Omega = 2 \omega_{jk}$. In this case,  by
using the properties (\ref{eq:uvnorm1})-(\ref{eq:uvnorm2}), it is
also possible to show that ${\rm Im}[\Gamma_{+-}^{jj}]=0$  and
$\Gamma_{++}^{jj}=0$.

These are the main results of the two-mode approximation.  We note
that the evaluation of the growth rate (\ref{gamma2Mapp}) only
requires the solution of the Bogoliubov equations (\ref{eq:BOGO})
at $s_0$ yielding the frequencies $\omega_{jk},
\omega_{j'k}$ and the coupling element $\Gamma_{+-}^{jj'}$.

A further advantage of the two-mode approximation is that it
allows one to characterize the role of the seed excitations.
Let us concentrate for simplicity on the case of modes satisfying
the resonant condition $\Omega=\omega_{jk}+
\omega_{j'k}$. The solution of Eq.~(\ref{eq:c+-1eq})
with initial conditions (\ref{eq:ci}) has the form
\begin{equation}
f_{jk}(t)=\eta e^{\gamma_{\rm max} t}+\eta' e^{-\gamma_{\rm max}
t}\;, \label{eq:apprsol}
\end{equation}
where
\begin{eqnarray}
\eta &=& \frac{1}{2}[c_{jk}(0)-c_{j',-k}^*(0)e^{i\phi}]
\label{eq:eta}\\
\eta' &=& \frac{1}{2}[c_{jk}(0)+c_{j',-k}^*(0)e^{i\phi}]\;
\label{eq:eta1}
\end{eqnarray}
and $\phi=\mbox{phase}(\overline\gamma)$. For long time
($t\gg\gamma_{\rm max}^{-1}$),  the second term in
(\ref{eq:apprsol}) vanishes and the population of both resonant
modes have the same exponential growth with rate $\gamma_{\rm
max}$. The coefficient $\eta$ is related to the initial population
of the Bogoliubov modes (seed excitations) through
Eq.~(\ref{eq:eta}). More precisely, the parametric resonance
amplifies a quadrature of Bogoliubov excitations which is phase
locked with the perturbation \cite{walls2,phase}, and $\eta$ plays
the role of an effective seed. It is worth stressing that
exponential growth, i.e., the parametric instability, only occurs
if the Bogoliubov modes are already populated at the initial time
($\eta \neq 0$). In section \ref{section-seed} we will discuss how
this can be related to thermal and/or quantum fluctuations.

%%%%%%%%%%%%%%%%%%%%%%%%%%%%%%%%%%%%%%%%%%%%%%%%%%%%%%%%%%%%%%%%%%%%%%%%%%%%
\section{Tight binding regime}
\label{section-tb}

As the lattice is made deeper one enters the tight binding regime
where the condensate atoms are strongly localized near the minima
of the lattice potential and only next neighbor Wannier functions
have nonvanishing overlap. In this regime, one can derive an
analytic expression for the lowest Bogoliubov band (see for
example \cite{epjd})
\begin{equation}
\hbar\omega_0=\sqrt{\varepsilon_0\left(\varepsilon_0+2\kappa^{-1}
\right)}\,,
\label{eq:omega0tb}
\end{equation}
with $\varepsilon_0=2\delta\sin(kd/2)^2$. The tunnelling parameter
$\delta$ is related to the effective mass through the relation
$\delta=(2/\pi^2)(m/m^*)E_R$, while
$\kappa^{-1}=n\partial\mu/\partial n$ is the inverse of the
compressibility. Furthermore, as discussed in \cite{epjd}, the
Bogoliubov amplitudes of the lowest band become proportional to
stationary Bloch state solutions $\varphi_k$ of the GP equation,
$u_{0k}=U_k\varphi_k$, $v_{0k}=V_k\varphi_k$ with
\begin{eqnarray}
U_k = \frac{\varepsilon_0+\hbar\omega_0}{2
\sqrt{\hbar\omega_0\varepsilon_0}}\,,
\label{eq:Uk}\\
V_k = \frac{\varepsilon_0-\hbar\omega_0}{2
\sqrt{\hbar\omega_0\varepsilon_0}}
\label{eq:Vk}\,.
\end{eqnarray}
One can use these results to calculate the coupling element
$\Gamma_{+-}^{00}$ in Eq.(\ref{eq:Gamma+-jj'}). The result is
\begin{equation}
\Gamma_{+-}^{00}= U_k\partial_s V_k-V_k\partial_s U_k \,.
\label{eq:Gamma+-tbUV}
\end{equation}
Inserting Eqs.~(\ref{eq:Uk})-(\ref{eq:Vk}), one obtains
\begin{eqnarray}
\Gamma_{+-}^{00} & = &
\frac{\omega_0\partial_s\varepsilon_0
-\varepsilon_0\partial_s\omega_0}{2\varepsilon\omega} \nonumber \\
& = & \frac{\kappa^{-1}\partial_s\delta-
\delta\partial_s\kappa^{-1}}{2 \delta 
\left(\varepsilon_0+2\kappa^{-1}\right)}\, .
\label{eq:Gamma+-tb}
\end{eqnarray}
In the tight binding regime the coupling $\Gamma_{+-}^{00}$ is hence 
directly related to the lattice depth dependence of compressibility 
and tunnelling (effective mass). Expression (\ref{eq:Gamma+-tb}) 
allows for a comparison with the two mode approximation, as we will 
see in the next section. 

A nice result is obtained by inserting expression 
(\ref{eq:Gamma+-tb}) into Eq.~(\ref{gammamax2M}), using the 
definition (\ref{gammamax}) and taking the $s_0 \gg 1$ limit. 
One finds 
\begin{equation}
\gamma_{\rm max}=\frac{A s_0\Omega }{4c_{\rm hd}}
\left.\frac{\partial c_{\rm hd}}{\partial s}
\right|_{s=s_0}\;,
\label{gammaHD}
\end{equation}
with 
\begin{equation}
c_{\rm hd}= (m^* \kappa)^{-1/2} \,.
\label{cHD}
\end{equation}
The latter coincides with the Bogoliubov sound velocity of the 
hydrodynamic approach of Ref.~\cite{epjd}. Therefore, in the 
limit of very large $s_0$, where the order parameter $\Psi_0$
can be assumed to be time-independent and the fluctuating part
$\delta \Psi$ can be described in terms of hydrodynamic 
phonons with linear dispersion $\omega_{0k}= c_{\rm hd} k$, the
growth rate of the parametric resonance turns out to be 
determined by the change in sound velocity induced by the lattice 
modulation.

%%%%%%%%%%%%%%%%%%%%%%%%%%%%%%%%%%%%%%%%%%%%%%%%%%%%%%%%%%%%%%%%%%%%%%%%

\section{Results}
\label{section-results}

In this section, we present the results of the stability analysis
of section \ref{section-bogo} together with the ones of the two-mode
and tight-binding approximations of sections \ref{section-2mode} and
\ref{section-tb}.

\begin{figure}[h]
\begin{center}
\includegraphics[width=8cm]{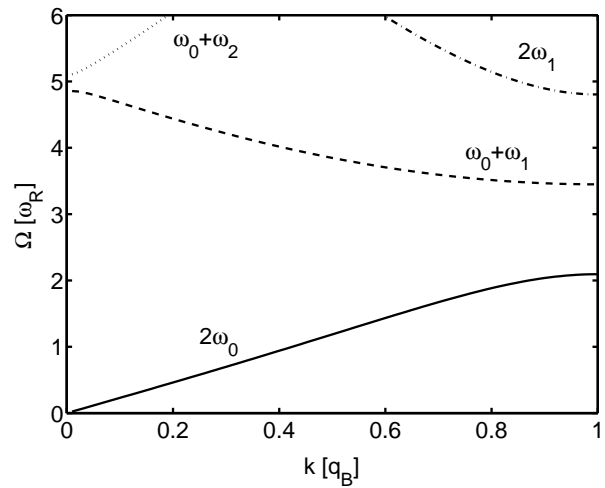}
\end{center}
\caption{Position of parametric resonances for $s_0=4, 
gn=0.72 E_R$ and for a small modulation amplitude, $A \to 0$. 
See Fig.\ref{QP0} for the Bogoliubov spectrum $\omega_{jk}$ with 
the same parameters.}
\label{QP1}
\end{figure}

Let us first discuss the limit of a small modulation amplitude,
$A \to 0$. In this limit the results of the stability analysis of
Eq.~(\ref{eq:rescaledBOGO}) are found to exactly coincide with the ones 
of the two-mode approximation. The parametrically unstable regions in
the $(\Omega,k)$ plane have a vanishingly small width, falling onto the 
lines where the resonance condition (\ref{eq:res-cond}) is 
fulfilled. The lowest lines are plotted in Fig.~\ref{QP1}. The 
different line styles indicate different pairs of resonant modes. 
By direct comparison with the spectrum in Fig.~\ref{QP0}, one sees 
that the first resonance is encountered at twice the frequency of 
the lowest Bogoliubov band, $\Omega=2\omega_{0k}$ (solid line). The 
next resonance is found when $\Omega$ is the sum of the frequencies 
of the lowest and the next Bogoliubov band, $\Omega=\omega_{0k}+
\omega_{1k}$ (dashed line). Further resonances occur at twice the 
frequencies of the first excited band, $\Omega=2\omega_{1k}$ (dash-dotted 
line), at the sum of frequencies of the lowest and the second excited 
bands, $\Omega=\omega_{0k}+\omega_{2k}$ (dotted line), and so on. One 
or more parametric excitations occur for any value of $\Omega$, except 
within the gap between $2\omega_0$ and $(\omega_0+\omega_1)$ resonances 
at zone boundary.

%%%%%%%%%%%%%%
%%%%%%%%%%%%%%

\begin{figure}[h]
\begin{center}
\includegraphics[width=8cm]{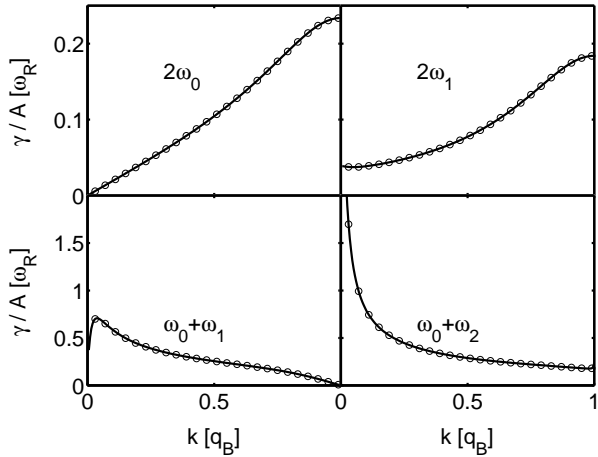}
\end{center}
\caption{Growth rates of parametric amplification for $s_0=4, 
gn=0.72 E_R$ in the small $A$ limit. The resonances are the same 
as in Fig.\ref{QP1}.  Lines: results of the stability analysis of 
Eq.~(\ref{eq:rescaledBOGO}). Different values of $A$ in 
the range $A<0.01$ give the same curves for $\gamma/A$. Circles: 
Eqs.~(\ref{gammamax2M}) and (\ref{gammamax}) of the two-mode 
approximation, with $\Gamma_{+-}^{jj'}$ given in (\ref{eq:Gamma+-jj'}) 
and calculated by solving the Bogoliubov equations (\ref{eq:BOGO}) 
at $s_0=4$. }
\label{QP2}
\end{figure}

For small $A$, both the growth rate of the unstable modes and the
width of the instability regions are found to be proportional to $A$.
In Fig.~\ref{QP2} we plot the growth rate on resonance, $\gamma =
\gamma_{\rm max}$, divided by $A$, for the same resonances of 
Fig.~\ref{QP1}. Solid lines are the numerical results of the stability 
analysis of section \ref{section-multimode} (see definition 
(\ref{relambda})) for $A < 0.01$, while empty circles represent the 
results of the two-mode approximation (see Eqs.~(\ref{gammamax2M}) 
and (\ref{gammamax})).  The very good agreement between the exact 
multi-mode calculations and the two-mode results shows that, for 
small $A$, the essential mechanism underlying the parametric 
process is fully captured by the two-mode approximation. In 
Fig.~\ref{QP2} the quantity $\gamma/A$ takes values ranging from 
$0$ to typically $\sim 0.2\omega_R$, which implies that, on resonance, 
parametric amplification occurs on a time scale much longer than 
$\omega_R^{-1}$.

%%%%%%%%%%%%%%
%%%%%%%%%%%%%%

Within the two-mode approximation, two excitations of frequency
$\omega_{ik}$ and $\omega_{jk}$ are unstable when $| \omega_{ik} 
+  \omega_{jk} - \Omega| < 2 |\overline{\gamma}|$, where 
$|\overline{\gamma}|=\gamma_{\rm max}$ is the maximal growth rate 
on resonance. The values of the maximal growth rate plotted in 
Fig.~\ref{QP2} thus indicate that, at small values of $A$, one 
is generally dealing with very narrow resonances. For the parameters 
used in Figs.~\ref{QP1} and \ref{QP2} ($s_0=4$ and $gn=0.72 E_R$) 
and for $A < 0.01$, the width is less than $1\%$ of the resonance 
frequency.

%%%%%%%%%%%%%%
\begin{figure}[h]
\begin{center}
\includegraphics[width=8cm]{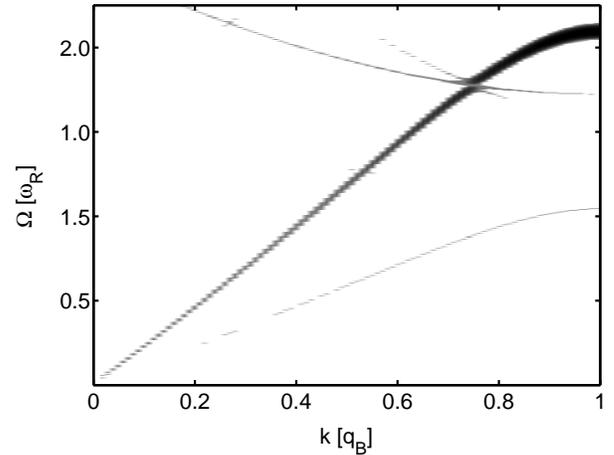}
\end{center}
\caption{Stability diagram in the
$(k,\Omega )$-plane obtained with the procedure of section
\protect\ref{section-multimode} for $A=0.1$, $s_0=4$ and
$gn = 0.72E_R$. The lightest level of grey in the grey-scale
represents all modes having $10^{-10} < \gamma < 10^{-3}$. The
main branch corresponds to the $2 \omega_0$-resonance, whose 
maximal growth rate $\gamma_{\rm max}$ grows with $k$ as 
shown in Fig.~\protect\ref{QP4}.  The other weak branches
are higher order parametric resonances. 
}
\label{QP3}
\end{figure}
%%%%%%%%%%%%%%

%%%%%%%%%%%%%%
\begin{figure}[h]
\begin{center}
\includegraphics[width=8cm]{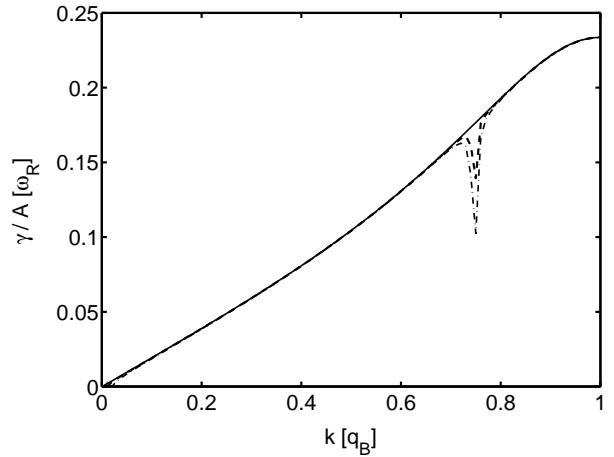}
\end{center}
\caption{Growth rate $\gamma/A$ for the $2\omega_0$- resonance 
($\Omega=2\omega_{0k}$) obtained with 
the stability analysis of Eq.~(\ref{eq:rescaledBOGO}) with $s_0=4$, 
$gn=0.72E_R$ and $A<0.01$ (solid line) and $A=0.05$ (dashed) 
and $0.1$ (dot-dashed). }
\label{QP4}
\end{figure}
%%%%%%%%%%%%%%

The overall qualitative picture is preserved when the modulation
amplitude $A$ is tuned to experimentally relevant values. An example
is given in Fig.~\ref{QP3} where we plot the stability diagram in
the energy range of the $2\omega_0$-resonance obtained with the
stability analysis of section \ref{section-multimode} for $A=0.1$.
The main branch remains centered around the resonance condition 
$\Omega = 2\omega_{0k}$. The shape of $\gamma(k,\Omega)$ near 
resonance is very well approximated by Eq.~(\ref{gamma2Mapp}),
which yields $\gamma= [ \gamma_{\rm max}^2 - \left(\omega_{0k} - 
\Omega/2 \right)^2 ]^{1/2}$. The maximal growth rate $\gamma_{\rm max}$ 
is plotted as a dot-dashed line in Fig.~\ref{QP4}. Figure~\ref{QP3}
shows that the $2\omega_0$-resonance remains narrow even for relatively
large $A$, the finite width becoming noticeable only close to the 
band edge where it reaches about $5$\% of the resonance frequency. 
However, new features appear, 
which are due to very weak resonances of higher order in $A$. In 
particular, one clearly sees narrow unstable regions at frequency 
$\omega_0$ and $(\omega_0+\omega_1)/2$. The growth rate and the 
width of these resonances are both of order $A^2$, so that they are 
not accounted for by the two-mode approximation of section 
\ref{section-2mode}. The resonances at $(\omega_0+\omega_1)/2$ and 
$2 \omega_0$ exhibit a crossing, giving rise to a complex structure 
in the stability diagram.  For our values of $s_0$ and $gn$, the
crossing occurs at about $k=0.75 q_B$. The effects of this 
crossing are also visible in the growth rate of the $2\omega_0$ 
resonance. In Fig.~\ref{QP4} we plot $\gamma/A$ calculated at 
$\Omega = 2 \omega_{0k}$ for different values of $A$. The 
solid line is the same as in the top-left panel of Fig.~\ref{QP2} and 
corresponds to $A<0.01$. The dashed and dot-dashed lines correspond
to $A=0.05$ and $0.1$, respectively. The figure confirms that $\gamma$
remains directly proportional to $A$ up to relatively large values of 
$A$, except in a narrow region near $k\sim 0.75 q_B$, in correspondence 
with the crossing point. 

%%%%%%%%%%%%%%
\begin{figure}[h!]
\begin{center}
\includegraphics[width=8cm]{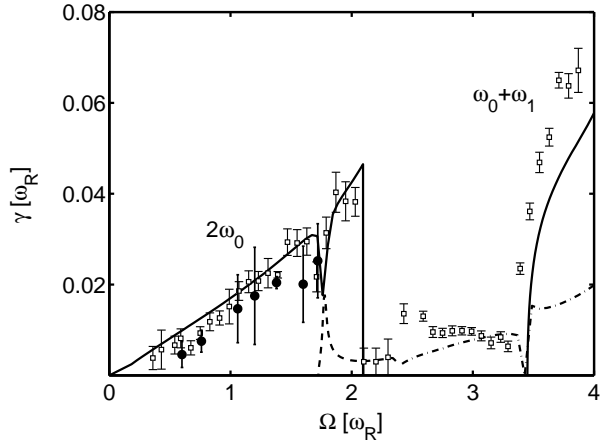}
\end{center}
\caption{Growth rate $\gamma$ as a function of the modulation 
frequency $\Omega$. The lines are the results of the stability analysis
of section \ref{section-bogo} for $A=0.2$, $s_0=4$ and $gn=0.72 E_R$.
In particular, the two solid lines are the resonances of order $A$
at $2\omega_0$ and $\omega_0+\omega_1$, while the dashed and 
dot-dashed lines are the weaker resonances of order $A^2$ at 
$(\omega_0+ \omega_1)/2$ and $(\omega_0+\omega_2)/2$, respectively. 
Empty circles are extracted from the time evolution of the incoherent 
fraction $\delta N/N$ in GP simulations on a cylindrical condensate, 
unbound along $z$ and harmonically confined in the transverse 
direction. Solid circles come from GP simulations as well, but for 
a condensate which is harmonically confined also along $z$, similar 
to the trapped condensates of the experiments of 
Ref.~\protect\cite{stoeferle,koehl,schori}. 
}
\label{GP2}
\end{figure}
%%%%%%%%%%%%%%

The growth rate can also be plotted as a function of the modulation 
frequency $\Omega$, as done in Fig.~\ref{GP2} where we show the 
results for $A=0.2$. The solid lines represent the growth rate
of the two resonances, of order $A$, at $2\omega_0$ and $\omega_0
+\omega_1$, while the dashed and dot-dashed lines correspond to the 
weaker resonances, of order $A^2$, at $(\omega_0+\omega_1)/2$ and 
$(\omega_0+\omega_2)/2$, respectively. The growth rate of 
the $\omega_0$ resonance is very small on this scale and it is 
not shown. The points with error bars come from GP simulations 
and will be discussed in the next section. 
    
\begin{figure}[h!]
\begin{center}
\includegraphics[width=7cm]{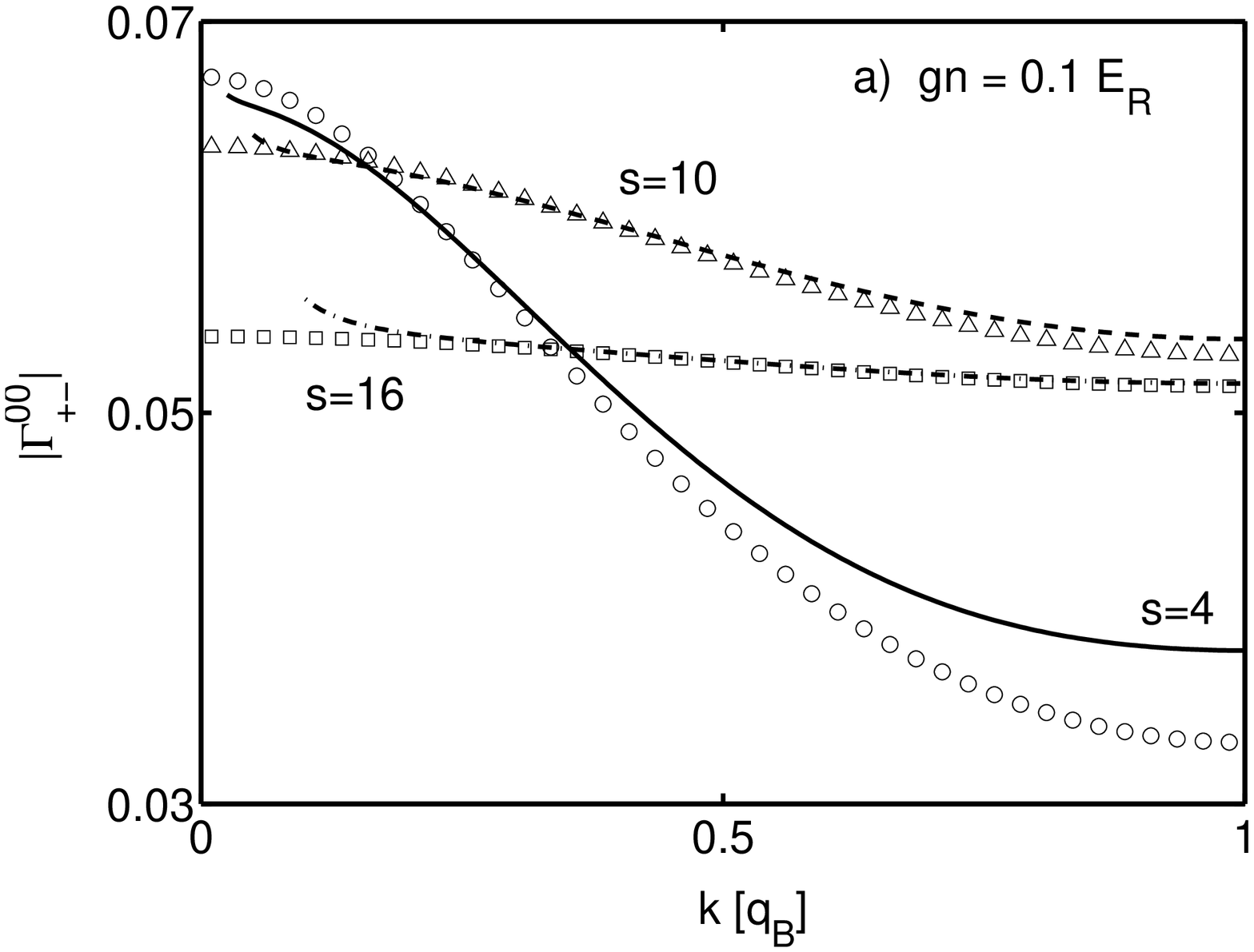} \\
\includegraphics[width=7cm]{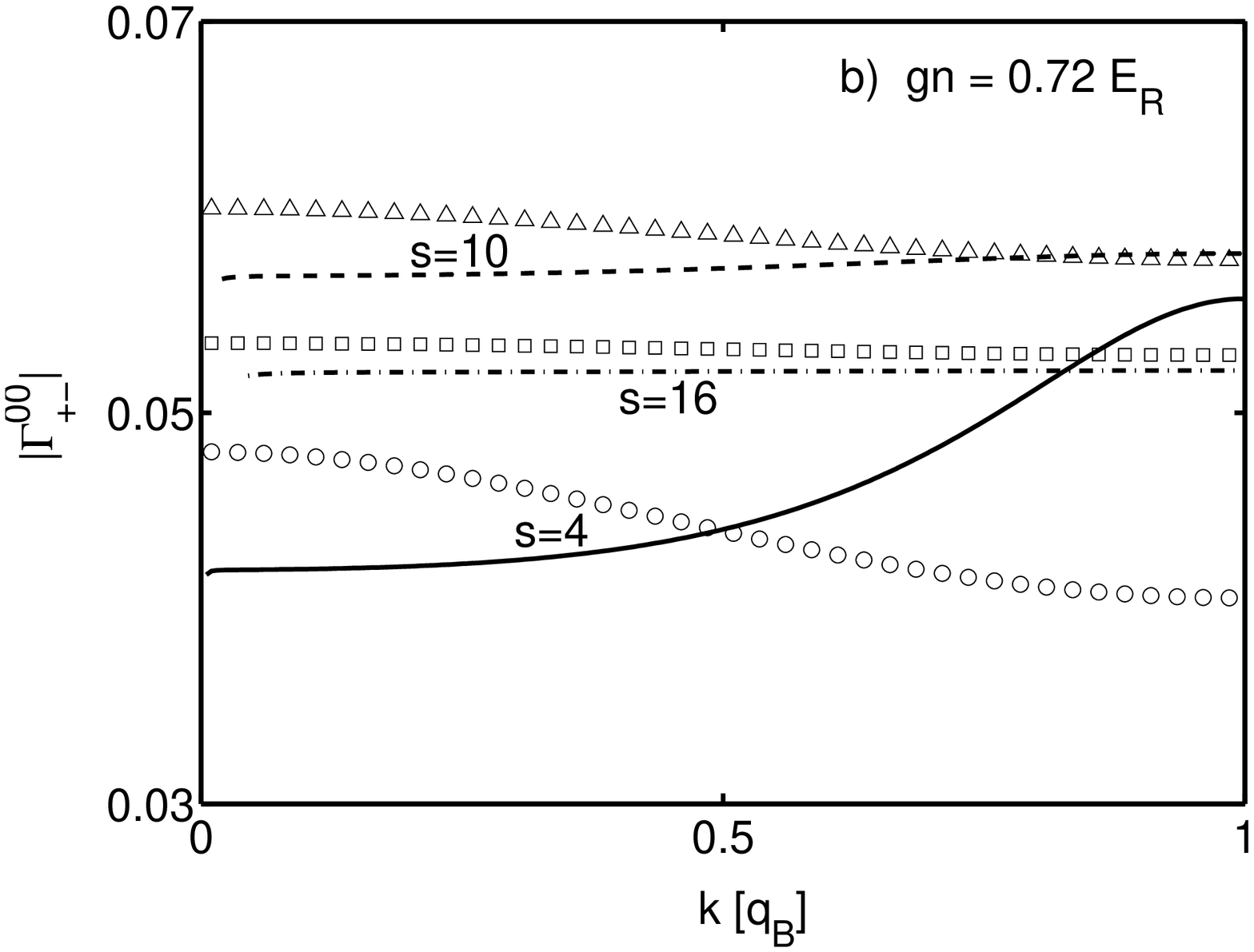}
\end{center}
\caption{Coupling $\Gamma_{+-}^{00}$ between counter-propagating
modes of the first Bogoliubov band as a function of quasimomentum 
$k$ at $gn=0.1E_R$ (top panel) and $gn=0.72E_R$ (bottom panel) for 
different values of the lattice depth $s_0=4, 10$ and $16$. Lines: 
definition (\protect\ref{eq:Gamma+-jj'}) with the quasiparticle 
amplitudes taken from the numerical solution of the Bogoliubov 
equations (\ref{eq:BOGO}). Points: Tight binding result 
(\ref{eq:Gamma+-tbUV}).}
\label{QP7}
\end{figure}

Coming back to the $A\to 0$ limit, we notice that, as shown in
Fig.~\ref{QP2},  the growth rates of the various types
of resonances significantly differ in their dependence on the
quasimomentum $k$.  Apart from a slight upward bend at around
half the Brillouin zone, the curve for the $2\omega_0$-resonance
mirrors the form of the lowest Bogoliubov band (see Fig.~\ref{QP0}).
In the other cases, the $k$-dependence of the growth rates is not
that easily characterized.  Some insight is gained from the two-mode
approximation where, according to expression (\ref{gamma2Mapp}), the
$k$-dependence of the growth rate on resonance is determined by the
product $(\omega_{jk}+\omega_{j'k})\Gamma_{+-}^{jj'}(k)$. Hence, the
fact that for the $2\omega_0$-resonance the growth rate $\gamma$
has a shape similar to that of $\omega_{0k}$ implies that 
$\Gamma_{+-}^{00}$ varies little with $k$. This is confirmed by a 
direct calculation of $\Gamma_{+-}^{00}$ through 
Eq.~(\ref{eq:Gamma+-jj'}). The result is shown as the solid line 
in the lower panel of Fig.~\ref{QP7}. Conversely, the growth rates 
of the other resonances in Fig.~\ref{QP2} do not exhibit this 
behavior, indicating that the corresponding $\Gamma_{+-}^{jj'}$ 
have a nontrivial dependence on $k$.

%%%%%%%%%%%%%%
%%%%%%%%%%%%%%

Up to now, our discussion has dealt with the properties of
parametric instabilities at a lattice depth which is modulated
around $s_0=4$. As $s_0$ is increased, the width of the first
Bogoliubov band decreases and, accordingly, the $2\omega_0$
resonance is shifted to lower modulation frequencies $\Omega$.
Moreover, an increase of the lattice depth changes the growth
rates of parametric amplification. This can be seen in the two-mode
approximation where the $2\omega_0$-growth rate is proportional
to $s_0$, the Bogoliubov frequency $\omega_0$ and the coupling
$\Gamma_{+-}^{00}$. In the lower panel of Fig.~\ref{QP7} the
quantity $\Gamma_{+-}^{00}$ is plotted for three different values
of $s$. As one can see, its $s$ dependence is quite weak. The
product $s_0\omega_0$ remains also of the same order, since an
increase of $s_0$ is counterbalanced by a decrease of $\omega_0$.
At zone boundary and for $gn=0$, their product takes
the values $4.2, 5.8, 5.2 E_R$ for $s_0=4, 10, 16$, respectively.
This implies that the effects of a variation of $s_0$ on the
timescale of parametric instability of the $2 \omega_0$ branch
are not dramatic in this range of $s_0$. For sufficiently large
$s_0$ one enters the tight binding regime, where $\omega_0$ is
proportional to $\sqrt{\delta \kappa^{-1}}$ (see
Eq.(\ref{eq:omega0tb})) and one gets an exponential decrease 
which eventually dominates the lattice depth dependence of the 
product $s_0\omega_0$ and hence of the growth rate.

In section \ref{section-tb} above, we derived an expression for the
coupling $\Gamma_{+-}^{00}$ in the tight binding regime (see
Eqs.(\ref{eq:Gamma+-tbUV})-(\ref{eq:Gamma+-tb})). In Fig.~\ref{QP7}
we compare the tight binding results with those obtained
from the numerical solution of the Bogoliubov equations
(\ref{eq:BOGO}) for different values of the lattice depth $s_0$ and
two different values of the interaction parameter $gn$. The
numerical results (lines) approach the tight binding results (points)
as the lattice is made deeper. For a given $s_0$ the agreement is
better when the interaction parameter $gn$ is lower. For $gn=0.1E_R$
the agreement is reasonably good already at a lattice depth of $s_0=4$,
while deeper lattices are necessary to achieve the same degree of
agreement for $gn=0.72E_R$. This is consistent with general arguments
on the applicability of the tight binding approximation \cite{epjd}.
A repulsive mean-field interaction causes an effective smoothing of
the periodic potential seen by the atoms and therefore the convergence
to the tight binding regime is pushed to higher values $s_0$.

With reference to Fig.~\ref{QP7} we note that, at fixed $s_0$, the
coupling $\Gamma_{+-}^{00}$ takes similar values for the two different
interaction parameters $gn$. However, as pointed out in section
\ref{section-mod}, the coupling between counter-propagating Bogoliubov
modes, which is necessary for parametric amplification, vanishes 
in the noninteracting case.  Numerically we find that the parameter
$gn$ has to be tuned to very small values ($\ll 10^{-3}$) before
a significant reduction of $\Gamma_{+-}^{00}$ is achieved.

%%%%%%%%%%%%%%%%%%%%%%%%%%%%%%%%%%%%%%%%%%%%%%%%%%%%%%%%%%%%%%%%%%%%%%%%%%
\section{Comparison with GP simulations}
\label{section-gp}

The above discussion of the parametric instability of the
condensate is based on a suitable linearization of the 1D GP
equation (\ref{eq:TDGP}). We assumed that most of the atoms are
described by an order parameter $\Psi_0$, which adiabatically
follows the modulations of the lattice depth $s(t)$, plus a small
deviation $\delta \Psi$. Under this hypothesis, we have found that
$\delta \Psi$ can exhibit an exponential growth as a consequence
of the parametric instability of Bogoliubov excitations. The
applicability of this approach to actual 3D condensates in
experimentally feasible conditions can be tested by comparing its
results with those of GP simulations, i.e., a direct numerical
integration of the time-dependent GP equation (\ref{tdgpe}). 

For condensates subject to strong transverse confinement the 
integration of the GP equation can be efficiently performed by 
using the Non-Polynomial Schr\"odinger Equation (NPSE) introduced 
in Ref.~\cite{npse}. The key assumption is that the order parameter
can be factorized, as in section \ref{section-bogo}, in the product of 
a Gaussian radial component and an axial wave function
$\Psi(z,t)$. Differently from section \ref{section-bogo}, using the
NPSE one assumes the Gaussian to have a $z$-and $t$-dependent 
width, $\sigma(z,t)$. The GP equation (\ref{tdgpe}) thus yields
\begin{equation}
i\hbar \partial_t \Psi =
\left[ -\frac{\hbar^2}{2m} \left( \partial_z^2
+\frac{1}{\sigma^2} +\frac{\sigma^2}{a_\perp^4} \right)
+ V +\frac{g_{3D}N|\Psi|^2}{2\pi\sigma^2} \right] \Psi \, ,
\label{npse}
\end{equation}
and $\sigma=a_{\perp}(1+2aN|\Psi|^2)^{1/4}$.  In the geometry of 
the present work, differences between NPSE and the exact 3D GP 
equation are negligible and solving the NPSE is much less time 
consuming. The effective 1D GP equation (\ref{eq:TDGP}) corresponds 
to the approximation $\sigma=a_\perp$. 

Let us first consider, as in section \ref{section-bogo}, a condensate 
which is unbound along $z$, subject to the external potential
\begin{equation}
V (z,t) = \frac{m}{2} \omega_{\perp}^2 r_\perp^2
          + s(t) E_R \sin^2 (q_B z)
          \, ,
\label{v}
\end{equation}
with $s(t)$ given in (\ref{eq:modulation}). In order to compare the
results with those of the effective 1D GP equation (\ref{eq:TDGP})
we choose the parameters in (\ref{npse}) and (\ref{v}) such that
the condensate has the same interaction parameter $gn$, where $n$
is the average linear density. 

\begin{figure}[h!]
\begin{center}
\includegraphics[width=8cm]{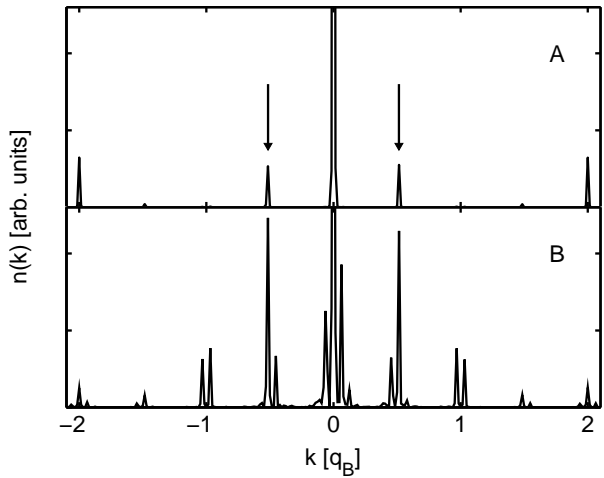}
\end{center}
\caption{Momentum distribution of a condensate which is
harmonically confined in the radial direction and periodic 
along $z$, obtained by performing GP simulations. The interaction 
parameter is $gn=0.72 E_R$, where $n$ is the average linear 
density. The optical lattice has $s_0=4$ and is modulated with 
$A=0.2$ and $\Omega=\omega_R$. The modulation starts at $t=0$ 
and the momentum distribution is plotted at $t=t_A$ (top panel) 
and $t=t_B$ (bottom panel), where $A$ and $B$ are the points 
indicated in Fig.~\protect\ref{GP1}. The arrows show the position 
of  the parametrically unstable modes. } 
\label{GP3}
\end{figure}

We first calculate the ground state order parameter $\Psi_0$ in a
static lattice of depth $s_0$ and then we follow the evolution of
$\Psi$ under the modulation $s(t)$, with given $A$ and $\Omega$. The
Fourier transform of $\Psi(z,t)$ gives the momentum distribution,
$n(k)$, as a function of time. The initial momentum distribution 
is charaterized by the three peaks at $k=0$ and $k=\pm 2q_B$, 
associated with the stationary order parameter $\Psi_0$ in
the periodic lattice. As time goes on, we observe a relative
oscillation of these peaks with the same frequency of the
lattice modulation. This reflects the fact that $\Psi_0$ 
adiabatically follows the oscillations of the lattice depth. 
However, after a certain time a parametric instability becomes 
visible, that is, the momentum distribution develops components 
at the momenta $\pm k$ of the amplified Bogoliubov modes. An 
example is shown in Fig.~\ref{GP3}. For even longer times, 
contributions from momenta throughout the first Brillouin zone 
start showing up due to higher-order harmonic generation, Bragg 
reflection and nonlinear coupling between different modes 
(see \cite{kraemer} for a detailed discussion of the
different stages of the evolution). 

\begin{figure}[h!]
\begin{center}
\includegraphics[width=8cm]{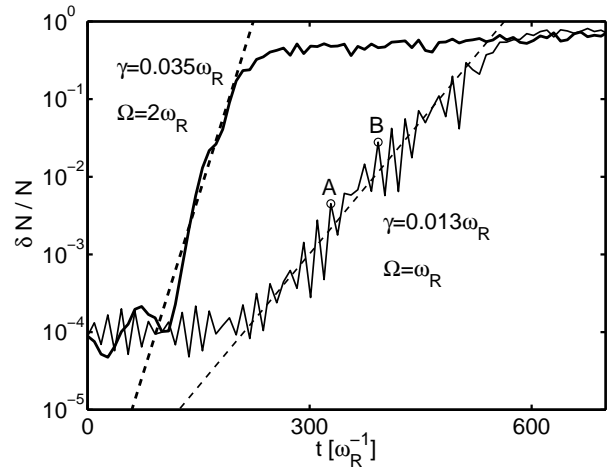}
\end{center}
\caption{Incoherent fraction $\delta N/N$ as a function of 
modulation time obtained by performing time-dependent GP 
simulations. Solid lines correspond to $s_0=4$, $A=0.2$ and 
two different modulation frequencies: $\Omega=\omega_R$ and 
$\Omega=2\omega_R$. The growth rate $\gamma$ is extracted from 
a linear fit (dashed lines). The momentum distribution of
the condensate at the instants $A$ and $B$ is shown in 
Fig.~\protect\ref{GP3}.} 
\label{GP1}
\end{figure}

The degree to which the parametric instability has evolved in 
time can be quantified by looking at the number of particles 
$\delta N$ contributing to the momentum distribution away from 
the $\Psi_0$ components. In practice, we calculate the {\it coherent} 
part, $N_{\rm coh}$, by integrating the momentum distribution within 
the intervals $0\pm \Delta$ and $\pm 2q_B\pm \Delta$, with $\Delta =
q_B/10$, and then we define the incoherent fraction $\delta N/N  =
(N-N_{\rm coh})/N$. In Fig.~\ref{GP1}, we plot $\delta N/N$ as a
function of time as obtained with a modulation amplitude $A=0.2$ for
two different values of the modulation frequency $\Omega$ within
the range of the $2\omega_0$-resonance.

%%%%%%%%%%%%%%
%%%%%%%%%%%%%%

In order to observe the amplification, an initial seed is
required. In the calculations a numerical noise is always present 
and is enough to trigger the instability with long modulation times. 
Adding an extra noise by hand in the initial order parameter 
anticipates the time at which the exponential growth becomes visible, 
but without affecting the growth rate.  For the calculations in
Figs.~\ref{GP3} and \ref{GP1} the seed is added in the form of 
a white numerical noise in  $\Psi(z,0)$. This noise, which is 
not visible on the scale of Fig.~\ref{GP3}, is added in the 
interval $-q_B < k <q_B$ and corresponds to $\delta N/N$ of the 
order of $10^{-4}$. Only a small part of this noise is 
subsequently amplified, i.e., the one having the quasimomentum 
$k$ of the resonant modes. By extrapolating the linear fit to 
$t=0$ we see that the seed in Fig.~\ref{GP1} is of the order 
of $10^{-7}$.

We extract the growth rates $\gamma$ by plotting the data on a
logarithmic scale and performing a linear fit in the time interval
where the exponential amplification is visible, as indicated by the
dashed lines in Fig.~\ref{GP1}. The extracted values of $\gamma$
can be directly compared to the results discussed in the previous
sections, which were calculated by an appropriate linearization of the
effective 1D GP equation (\ref{eq:TDGP}). The comparison is
shown in Fig.~\ref{GP2}. The results of the numerical integration of
Eq.~(\ref{tdgpe}) are shown as empty squares. The error bars come
from the uncertainties in the fitting procedure on the growth of
$\delta N/N$ in the simulations. The overall qualitative agreement 
is good. The growth rate increases almost linearly with $\Omega$,
following the $2 \omega_0$ resonance up to the band edge at $k=q_B$.
Then it suddenly drops. Weak $A^2$-resonances become visible in the 
gap between the $2\omega_0$ and the $(\omega_0+\omega_1)$ branches. 
The agreement is especially good for the lowest $2 \omega_0$ 
resonance. This confirms the accuracy, in this range of $\Omega$, of 
the approximations made in section \ref{section-bogo}, namely, that 
the 1D GP equation can be linearized in $\delta \Psi$ {\it via} the 
ansatz (\ref{deltapsi1}), with a $\Psi_0$ which adiabatically follows
the modulation of the lattice,  and that the inhomogeneous term
in (\ref{newTDGP}) is negligible.

%%%%%%%%%%%%%%
%%%%%%%%%%%%%%

Now we consider a condensate which is also axially trapped. We solve
again Eq.~(\ref{npse}), with the external potential
\begin{equation}
V (z,t) = \frac{m}{2} ( \omega_{\perp}^2 r_\perp^2 + \omega_z^2 z^2)
           + s(t) E_R \sin^2 (q_B z)         \,.
\label{vtrapped}
\end{equation}
We choose the parameters such as to simulate a single tube of
Ref.~\cite{stoeferle}. So, we use $d=413$ nm, $\omega_z=2\pi
\times 84.6$~Hz, $\omega_{\perp}=2 \pi \times 36.5$~KHz and the total
number of particles $N=100$. These values yield $E_R/\hbar = 2\pi
\times 3.34$ KHz. These are the same parameters used in our previous
work \cite{kraemer}. We extract the growth rate as done above for
the simulation with the infinite cylinder. The results for the
$2\omega_0$ resonance are shown in Fig.~\ref{GP2} as solid circles.
We notice that the number of particles per site in the trapped
condensate is $z$-dependent, so that the comparison with the 
infinite cylinder of fixed linear density $n$ is subject to some
arbitrariness in the definition of the axial average of $n$ in
the trap. Nevertheless the agreement is rather satisfactory and
supports the physical pictures discussed in  \cite{kraemer} about
the key role played by parametric resonances in the experimentally
observed response to a lattice modulation in the superfluid regime
\cite{stoeferle,koehl,schori}.

Finally, it is worth stressing that GP simulations account for the
effects of the nonlinear dynamics associated with the mean-field
interaction term in the time dependent GP equation. In particular,
when the population of a parametrically unstable mode becomes
large, nonlinear processes yield a significant coupling with other
nonresonant modes and a consequent redistribution of energy over a
wider range of quasimomenta, eventually causing a saturation of the
incoherent fraction $\delta N/N$ when it approaches $1$, as in the 
two curves plotted in Fig.~\ref{GP1}. Effects of these processes 
are also visible in part B of Fig.~\ref{GP3}. In the case of 
axially trapped condensates they are enhanced by the finite size of 
the systems, $Z$, which introduces a natural broadening of the order 
$1/Z$ in the momentum distribution (see discussion in \cite{kraemer}). 
However, an important result of the present analysis is that such 
nonlinear processes do not prevent the parametric resonances to be 
observable, either by measuring the final effects of the amplification 
process (as, for instance, the total energy transferred to the 
condensate as a function of $\Omega$ and $A$ as in 
\cite{stoeferle,koehl,schori}) or by looking directly at the 
exponential growth of the unstable modes in the momentum distribution. 
In the latter case, the identification of a peak with a rapidly 
growing height in the measured momentum distribution would also 
allow one to use the parametric resonances as a spectroscopy tool, 
namely to measure the dispersion relation $\omega_{jk}$ in a 
parameter space where other spectroscopic techniques, such as 
two-photon Bragg scattering, are not easily applicable. The timescale 
and the size of $\delta N/N$ predicted in this work seem well 
compatible with feasible experiments.

%%%%%%%%%%%%%%%%%%%%%%%%%%%%%%%%%%%%%%%%%%%%%%%%%%%%%%%%%%%%%%%%%%%%%%%%%%
\section{Remarks on thermal and quantum seed}
\label{section-seed}

As we already noticed, at the mean-field level the parametric 
resonances occur only if the relevant Bogoliubov modes are 
initially occupied. Within the two-mode approximation, discussed 
in section \ref{section-2mode}, we have shown that a special 
role is played by the effective seed $\eta$ defined in (\ref{eq:eta}). 
This quantity is exponentially amplified in time.

Different mechanisms can cause a non-vanishing $\eta$. The
condensate can be simply out of equilibrium at the beginning of
the modulation due to non-adiabatic processes when it is loaded in
the trap and in the optical lattices. This yields some excitations
over the groundstate. A similar and controllable seed
could be added to the condensate by exciting certain Bogoliubov
modes by means of an appropriate external perturbation (e.g.,
two-photon Bragg scattering) before starting the lattice
modulation. Both these types of seed can be described with the GP
equation. On the contrary, the GP equation does not account for
thermal and quantum fluctuations, which have to be treated with
different methods beyond mean-field. 

In sections \ref{section-bogo} and \ref{section-2mode} we treated the 
collective excitations of the condensate by linearizing the GP equation. 
We have also used the Bogoliubov quasiparticle amplitudes, solution 
of Eq.~(\ref{eq:BOGO}), as basis functions to represent the fluctuating 
component $\delta \Psi$, the coefficients of the expansion 
(\ref{atilde}) being c-numbers related to the population of
the collective modes. This allowed us to interpret the dynamics in 
terms of coupling and growth of Bogoliubov excitations. A natural 
generalization consists in replacing these c-numbers with the
quasiparticle creation and annihilation operators, as in the
standard Bogoliubov theory \cite{bogo,kagan2}. In such a way one can 
treat both thermal and quantum fluctuations, under the only 
assumption that the total number of particles $N$ is macroscopic 
and the fraction of noncondensed particles is much smaller than 
$N$. A direct connection between this quantum treatment of the
operator $\delta \hat\Psi$ and our previous approach of 
section \ref{section-2mode} can be found by using the Wigner 
representation of the quantum fields \cite{sinatra,walls}. In 
this way, the dynamics is still governed by the classical equations 
of sections \ref{section-bogo} and \ref{section-2mode}, but the 
condensate depletion is now included {\it via} a stochastic 
distribution of the coefficients $c_{jk}$. Exact results can be 
obtained by averaging over many different realizations of the 
condensate in the same equilibrium conditions. In particular, one has 
\begin{eqnarray}
\langle c_{jk}(0) \rangle = 0 \; \; \; \; &,& \; \; \; \; \langle
c_{jk}(0) c_{j'k}(0) \rangle  = 0 \; ,
\label{eq:Wigner0}\\
\langle c_{jk}(0)c^*_{j'k'}(0) \rangle &=&
\left[\frac{1}{2}+\frac{1}{e^{\beta \hbar\omega_{jk}} -1}\right]
\delta_{jj'} \delta_{kk'} \; , 
\label{eq:Wigner}
\end{eqnarray}
where $\beta = (k_B T)^{-1}$. 
The $1/2$ term in (\ref{eq:Wigner}) describes quantum fluctuations
and is the leading one at small $T$, while the second term
accounts for thermal fluctuations. One can make use of
(\ref{eq:Wigner0})-(\ref{eq:Wigner}) to evaluate the mean
effective seed defined in (\ref{eq:eta}), finding 
\begin{eqnarray}
\langle |\eta|^2 \rangle &=& \frac{1}{4}\left(1+ \frac{1}{e^{\beta
\hbar\omega_{jk}}-1}+ \frac{1}{e^{\beta
\hbar\omega_{j'k}}-1}\right) .
\label{eq:etaT}
\end{eqnarray}
Equation (\ref{eq:etaT}) provides a way to extract information 
about the initial seed from the measurement of the exponential 
growth of the incoherent fraction.  For example, within 
the two mode approximation described in section \ref{section-2mode}, 
one finds that the number of particles in the modes $j$ and $j'$ at 
$t \gg \gamma_{\rm max}^{-1}$, averaged over many realizations, 
grows as 
\begin{equation}
\langle \delta N \rangle = C \langle |\eta|^2 \rangle 
e^{2\gamma_{\rm max} t} \; ,
\label{eq:deltaN}
\end{equation}
with 
$C=\int_{-d/2}^{d/2} dz \ 
(|\tilde u_{jk}|^2 + |\tilde v_{jk}|^2  + |\tilde u_{j'k}|^2 + 
|\tilde v_{j'k}|^2)$. So, if one is able to 
measure $\langle \delta N \rangle$ at different times during 
the parametric amplification and extract its value at $t=0$ 
from a fit, then Eqs.~(\ref{eq:etaT}) and (\ref{eq:deltaN})
give direct information on the quantum or thermal nature of 
the seed, provided the Bogoliubov spectrum of the condensate 
is known. 

The thermal regime is obtained when $k_B T \gg \hbar\omega_{jk}$ 
and corresponds to the ``classical'' limit where the occupation 
numbers of the relevant modes are large.  In this regime, the 
parametric response of the system to the lattice modulation can 
provide a thermometry. Let us suppose, in fact, that
an experiment is set up to observe the exponential growth at 
$\Omega = 2 \omega_{0k}$, as in the simulation
of Fig.~\ref{GP1}, and that the dispersion law $\omega_{0k}$ is
also known (either from calculations or by direct measurements of
$n(k)$).  Equation~(\ref{eq:etaT}) predicts
$\langle |\eta|^2 \rangle \simeq (1/2) k_B T/ \hbar \omega_{0k}$
for $k_B T \gg \hbar\omega_{0k}$, so that the knowledge of the
spectrum and the average seed vs. $k$ provides an estimate of the
temperature $T$. If the fit works well this means that the gas of
excitations is in thermal equilibrium. This is an important
result in view of the fact that the temperature we are speaking 
about can be much smaller than the temperatures which can be 
measured with current techniques, based on the detection of the
thermal cloud.  In the absence of a visible thermal cloud, 
parametric resonances appear to be an interesting method to 
make small thermal fluctuation detectable {\it via} a selective
amplification. 

The other important limiting case is that encountered when $k_B T
\ll \hbar\omega_{jk}$. In this condition the initial population 
of the unstable modes is mainly due to quantum fluctuations 
and $\langle |\eta|^2 \rangle \sim 1/4$. The modulation of the
lattice thus acts on the quantum vacuum, squeezing it and 
tranferring energy to the system. So, even at $T=0$, one can 
parametrically excite the condensate by amplification of the
vacuum fluctuations and this is a signature of its underlying 
quantum nature. This regime is analogous to the parametric down 
conversion, a well-known phenomenon in nonlinear quantum 
optics \cite{walls}. A conceptually interesting point is that 
the amplification of quantum fluctuations acts as a source of 
entangled quasiparticles, analoguous to parametric sources of 
entangled photon pairs in quantum optics \cite{kwiat}.  
A particularly appealing case is that of a modulation frequency 
$\Omega$ such as to resonantly excite two counter-propagating 
quasiparticles in different bands, $j$ and $j'$. In this case, the
quasiparticle pair is represented by the entangled state 
\begin{equation}
|\Psi\rangle \propto   |j,k\rangle |j',-k\rangle 
                    +  |j,-k\rangle |j',k\rangle \; . 
\end{equation}
A quite similar situation has been recently discussed for the 
generation of branch-entangled polariton pairs in microcavities 
through spontaneous interbranch parametric scattering \cite{ciuti}. 
It is finally worth mentioning that the creation of quasiparticle
pairs out of vacuum fluctuations can also be viewed as a 
manifestation of a dynamic Casimir effect: the environment in 
which quasiparticles live is periodically modulated in time and 
this modulation transforms virtual quasiparticles into real 
quasiparticles \cite{casimir}.

%%%%%%%%%%%%%%%%%%%%%%%%%%%%%%%%%%%%%%%%%%%%%%%%%%%%%%%%%%%%%%%%%%%%%%%%%%
\section{Conclusions and outlook}

In this work we have explored the origin and the effects of parametric
resonances in elongated Bose-Einstein condensates subject to a 1D
optical lattice whose intensity is periodically modulated in time. We
have used a suitable linearization of the time-dependent GP equation
to calculate the stability diagram and the growth rate of unstable
modes. We have shown that the main mechanism of instability is a
coupling between pairs of counter-propagating Bogoliubov
excitations. This coupling is caused by the modulation of the
background in which the excitations live. This picture emerges quite
clearly when a two-mode approximation is considered.  In this case,
one can derive semi-analytic results for the coupling between
Bogoliubov modes in different bands and discuss the convergence
to the tight binding regime. The results have been compared with those
of time-dependent GP simulations, for both condensates which are
infinite and trapped in the axial direction. The overall agreement
supports the interpretation of our previous GP simulations \cite{kraemer}, 
which were aimed at explaining the response to the lattice modulation 
observed in the experiments of Refs.~\cite{stoeferle,koehl,schori} in 
the superfluid regime. Our analysis also suggests the
possibility to perform new experiments in order to observe the effects
of the parametric resonances in a more controllable way as a tool for
a novel type of spectroscopy and, possibly, for a characterization of
quantum and/or classical seeds in actual condensates. Along this line,
our final discussion about the role of the seed and the applicability
of the Wigner representation is more intended to be a suggestion for 
future work rather than a quantitative analysis. This topic certainly 
deserves further investigations.
 
Our formalism can be naturally generalized to different types of
periodic modulations and different geometries. An interesting 
example is the periodic translation (shaking) of the optical lattice 
as in the recent experiment of Ref.~\cite{gemelke}. Another simple 
case is the modulation of the transverse trapping frequency in 
elongated condensates with or without optical lattice. Work in this
direction is in progress \cite{michele}.

%%%%%%%%%%%%%%%%%%%%%%%%%%%%%%%%%%%%%%%%%%%%%%%%%%%%%%%%%%%%%%%%%%%%%%%%%%
\acknowledgments

We are particularly indebted to I.~Carusotto for valuable 
suggestions and comments. Fruitful discussions with M.~Modugno 
and L.~Pitaevskii are also acknowledged. One of us (M.K.) thanks 
the DFG for support.

%%%%%%%%%%%%%%%%%%%%%%%%%%%%%%%%%%%%%%%%%%%%%%%%%%%%%%%%%%%%%%%%%%%%%%%%%%


\begin{thebibliography}{16}

\bibitem{landau} L.D.~Landau and E.M.~Lifshitz, {\it Mechanics}
(Oxford, Pergamon, 1973).

\bibitem{arnold} V.I. Arnold, {\it Mathematical Methods of Classical
Mechanics} (Springer-Verlag, Berlin, 1989).

\bibitem{castin} Y. Castin and R. Dum, Phys. Rev. Lett. {\bf 79},
3553 (1997).

\bibitem{kagan} Yu. Kagan and L.A. Maksimov, Phys. Rev. A {\bf
64}, 53610 (2001).

\bibitem{kevrekidis} P.G.~Kevrekidis, A.R.~Bishop, and K.O.~Rasmussen,
J. Low Temp. Phys. {\bf 120}, 205 (2000)

\bibitem{ripoll} J.J.~Garc\'\i a-Ripoll, V.M.~P\'erez-Garc\'\i a, and
P.~Torres, Phys. Rev. Lett. {\bf 83}, 1715 (1999).

\bibitem{staliunas} K.~Staliunas, S.~Longhi, and G.J.~de~Valc\'arcel,
Phys. Rev. Lett. {\bf 89}, 210406 (2002).

\bibitem{salasnich}  L.~Salasnich, A.~Parola, and L.~Reatto, J. Phys. B:
At. Mol. Opt. Phys. {\bf 35}, 3205 (2002)

\bibitem{salmond} G.L.~Salmond, C.A.~Holmes, and G.J.~Milburn, Phys.
Rev. A {\bf 65}, 033623 (2002).

\bibitem{harout} H.L.~Haroutyunyan and G.~Nienhuis, Phys. Rev. A
{\bf 70}, 063603 (2004).

\bibitem{rapti} Z.~Rapti, P.G.~Kevrekidis, A.~Smerzi, and A.R.~Bishop,
J. Phys. B: At. Mol. Opt. Phys {\bf 37}, S257 (2004).

\bibitem{rmp} F.~Dalfovo, S.~Giorgini, L.P.~Pitaevskii and S.~Stringari,
Rev. Mod. Phys. {\bf 71}, 463 (1999).

\bibitem{book} L. Pitaevskii and S. Stringari, {\it Bose-Einstein
condensation} (Clarendon Press, Oxford, 2003).

\bibitem{kraemer} M.~Kr\"amer, C. Tozzo, and F. Dalfovo, e-print
cond-mat/0410122, Phys. Rev. A, in press. 

\bibitem{stoeferle} T.~St\"oferle, H.~Moritz, C.~Schori, M.~K\"ohl,
and T.~Esslinger, Phys. Rev. Lett. {\bf 92}, 130403 (2004).

\bibitem{koehl} M.~K\"ohl, H.~Moritz, T.~St\"oferle, C.~Schori, and
T.~Esslinger, J. Low Temp. Phys. {\bf 138}, 635 (2005).

\bibitem{schori} C.~Schori, T.~St\"oferle, H.~Moritz, M.~K\"ohl, and
T.~Esslinger, Phys. Rev. Lett. {\bf 93}, 240402 (2004).

\bibitem{epjd} M.~Kr\"amer, C.~Menotti, L.~Pitaevskii, S.~Stringari,
Eur. Phys. J. D 27, 247 (2003).

\bibitem{berg} K. Berg-S\o rensen, K. M\o lmer, Phys. Rev. A {\bf 58},
1480 (1998).

\bibitem{chiofalo} M.L. Chiofalo, M. Polini, M.P. Tosi, Eur. Phys.
J. D {\bf 11}, 371 (2000).

\bibitem{ballagh} S.A.~Morgan, S.~Choi, K.~Burnett, and M.~Edwards,
Phys. Rev. A {\bf 57}, 3818 (1998);   P.B.~Blakie, R.J.~Ballagh,
and C.W.~Gardiner, Phys. Rev. A {\bf 65}, 033602 (2002).

\bibitem{tozzo} C.~Tozzo and F.~Dalfovo, Phys. Rev. A {\bf 69},
053606 (2004).

\bibitem{notegamma++} One can prove that 
$\partial_t \sum_{j} |c_{jk}|^2 = 0$, when $\Gamma_{+-}^{jj'}=0$, 
by using Eq.~(\protect\ref{c_jk}) and noticing that 
$\Gamma_{++}^{jj'} + (\Gamma_{++}^{j'j})^* = 0$ due to the 
ortho-normalization conditions (\protect\ref{eq:uvnorm1}).

\bibitem{shirts} R.B.~Shirts, ACM Transactions on Mathematical
Software {\bf 19}, 377 (1993).

\bibitem{floquet} W. Magnus and S. Winkler, {\it Hill's Equation}
(Dover, New York, 1979).

\bibitem{walls2} The analogous case in quantum optics is treated in
Ref.~\protect\cite{walls}, chapter 5.1.

\bibitem{phase} Note that, if the modulation 
(\protect\ref{eq:modulation}) is changed by a phase $\Delta \phi$, 
the results (\protect\ref{eq:eta})-(\protect\ref{eq:eta1}) will 
still apply, but with $\phi = {\rm phase} (\overline \gamma) + 
\Delta \phi$, which means that a different quadrature is amplified.

\bibitem{npse} L.~Salasnich. A.~Parola, and L.~Reatto, 
Phys. Rev. A {\bf 65}, 043614 (2002).

\bibitem{bogo}  N.N.~Bogoliubov, J. Phys (Moscow) {\bf 11}, 23 
(1947).

\bibitem{kagan2} This type of generalization has been used by 
Yu. Kagan and L.A. Maksimov, Laser Phys. {\bf 12}, 106 (2002), to 
study the damping of the radial breathing mode of an elongated 
condensate without optical lattice. Within their approach, the 
damping at zero temperature originates from the parametric 
amplification of longitudinal phonons out of quantum fluctuations.  

\bibitem{sinatra} A. Sinatra, Y. Castin, C. Lobo, J. Mod. Opt. 
{\bf 47}, 2629 (2000).

\bibitem{walls} D.F.~Walls and G.J.~Milburn, {\it Quantum Optics}, 
(Springer Verlag, Berlin, 1994).

\bibitem{kwiat} Pa.G.~Kwiat, K.~Mattle, H.~Weinfurter, A.~Zeilinger,
A.V.~Sergienko and Y.~Shih, Phys. Rev. Lett. {\bf 75}, 4337 (1995).

\bibitem{ciuti} C.~Ciuti, Phys. Rev. B {\bf 69}, 245304 (2004).

\bibitem{casimir} The analogous case of photons created in an 
oscillating cavity is treated, for instance, in A.~Lambrecht, 
M.-T.~Jaekel, and S.~Reynaud, Phys. Rev. Lett. {\bf 77}, 615 (1996). 
See also C.~Braggio {\it et al.}, quant-ph/0411085 and references 
therein. 

\bibitem{gemelke} N.~Gemelke, E.~Sarajlic, Y.~Bidel, S.~Hong, 
and S.~Chu, e-print cond-mat/0504311. The purpose of this work 
was the observation of a period-doubling instability analogous to
similar dynamical instabilities in a condensate moving in 
an optical lattice. In addition to the period-doubling instability,
an unexpected type of excitations was observed at higher
modulation frequency. We guess that this could be the 
result of the parametric resonance at the frequency $\omega_0+
\omega_1$, which is the lowest one allowed by the symmetry of 
the problem. 

\bibitem{michele} M.~Modugno, C.~Tozzo, and F.~Dalfovo, in 
preparation. 

\end{thebibliography}
\end{document}